\documentclass[a4paper,twoside,10pt]{article}
\usepackage{graphicx,publaob,url}

\def\papertype{Contributed paper}
\begin{document}

% Title
\title{THE MATLAS SURVEY OF FAINT OUTSKIRTS OF BRIGHT GALAXIES}

% Authors
\authors{M. B\'ILEK$^1$ \lowercase{and} P.-A. DUC$^1$}

% Addresses and e-mails
\address{$^1$Universit\'e de Strasbourg, CNRS, Observatoire astronomique de Strasbourg (ObAS), UMR 7550, 67000 Strasbourg, France}
\Email{bilek}{astro.unistra}{fr}

% Running titles
\markboth{DEEP IMAGING}{M. B\'ILEK and P.-A. DUC}

% Abstract
\abstract{Deep imaging, that is imaging capable of capturing very low surface brightness extended objects, is a quickly growing field of extragalactic astronomy. Not only can new types of faint objects be discovered, but deep images of bright galaxies are very valuable, too, since they reveal faint signs of past galaxy collisions, the tidal features. Such ``archeological'' record can be exploited for investigating how galaxies formed. In the MATLAS survey, we obtained extremely deep images of 177 nearby massive elliptical and lenticular galaxies using the 3.5m Canada-France-Hawaii Telescope. In our contribution, we will present the various types of objects and features seen in our images, for example   tidal features,  faint star-forming regions in otherwise quenched galaxies, or faint dust clouds in our own Galaxy. Finally, we will introduce the deep-imaging efforts at the Milankovi\'c Telescope.}

% Section and subsection
\section{Deep imaging in general}

\begin{figure}
	\includegraphics[width=\textwidth]{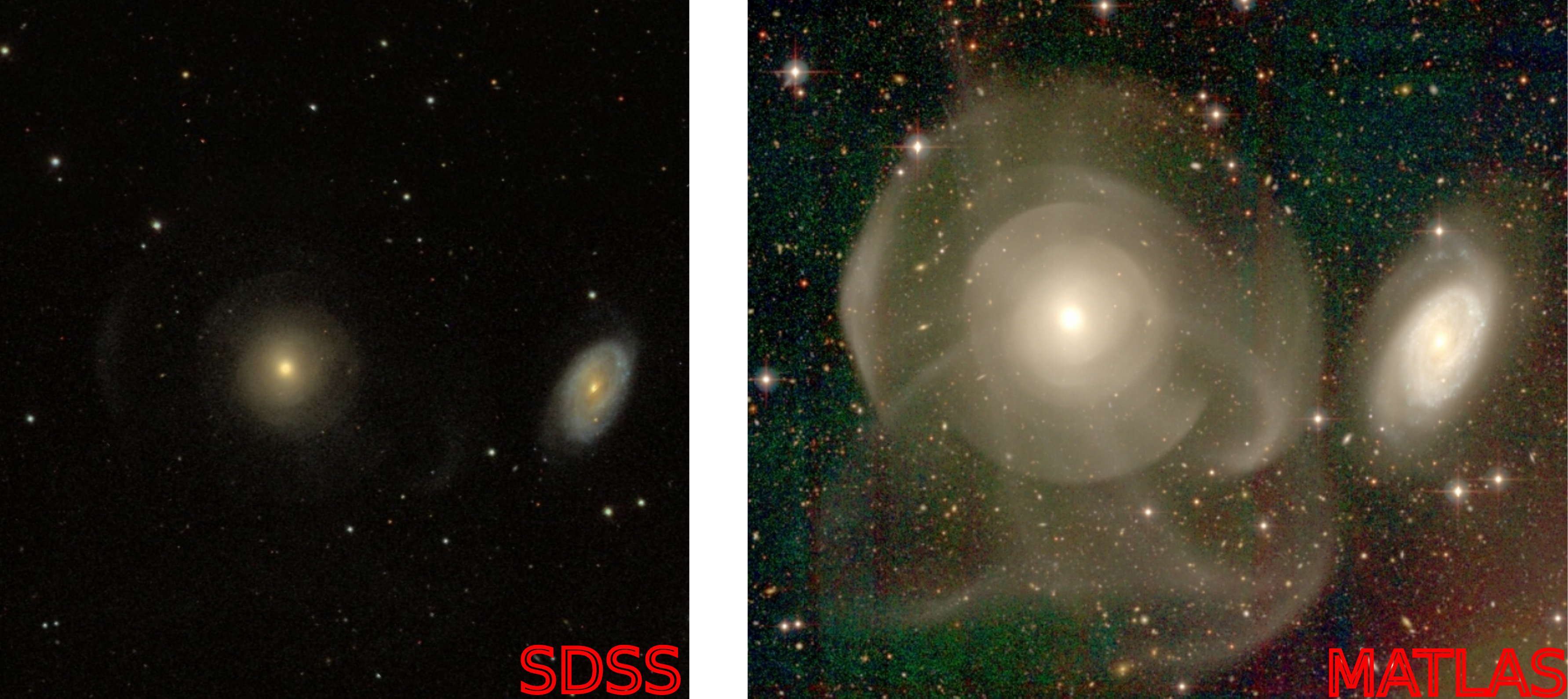}
	\caption{Illustration of deep imaging. Left panel: Standard image of a relatively bright and nearby galaxies NGC\,474 (left) and NGC\,470 (right) from SDSS. Right: Deep image of the same pair of  galaxies from the MATLAS survey shows the faint structures in the stellar halo with much better clarity.}
	\label{fig:comp}
\end{figure}

Much of our knowledge of galaxies comes from their images. Big sky surveys, such as the Sloan Digital Sky Survey (SDSS), took images of a huge number of galaxies. However, for the majority of galaxies, only images of their brightest parts are available. Figure~\ref{fig:comp}  shows an example. The left panel shows the galaxies NGC\,474 (left) and NGC\,470 (right) as imaged by SDSS. They are relatively bright, both having the apparent $V$-band magnitude of about 11.5. The image already reveals hints of the rich structure in the outer parts of   NGC\,474. The structures are seen with much better clarity  in the right panel. The image comes from the MATLAS  survey, that was optimized for detecting objects with a low surface brightness. This is very advantageous for investigating the formation of galaxies, as we will explain later. The survey was made possible because of the development of a technique called deep imaging.

The so called ``deep imaging'' technique presented here is a relatively new way of exploring the universe. Just as microscope enabled us to study objects with a small size or telescopes enabled us to explore the objects with a small angular size, deep imaging allowed us to explore the objects with a low surface brightness. But unlike in the previous two cases, the availability of new instruments has not been a request for the development of deep imaging. It could also be enabled because of new observing strategies and data processing techniques.  Obviously, the meaning of ``low surface brightness'' has evolved, but today we mean by that the value of about 28\,mag\,arcsec$^{-2}$. Reaching such a limit became routinely possible since about a decade even if the necessary telescopes and CCDs had been there before. Initially, the attempts of taking a deep image of an object were based just on collecting light for a long time at a fixed position or with restricted dithering. Then followed the standard calibration by bias, dark frame and flat field. But such an approach most often fails to reveal extended low-surface-brightness features. 

The faintest objects we are able to capture today have a surface brightness that is about 1000 times lower than the surface brightness of the sky at the darkest sites of the world. Separating such a small fraction of the useful signal from the total signal we receive is extremely sensitive to systematic errors. The most important systematic errors are the imperfections of the flat field images and the parasitic light, i.e. the sky light reflected from the mechanical parts of the telescope and camera on the CCD chip.  Important are also the errors coming from the variations of the sky brightness during the night and the scattered starlight (the wide wings of the point spread function). The standard calibration typically leaves artifacts that appear as large-scale brightness variations of the background (see the example in Duc et al. (2015). They are not too harmful as long as we are interested in small objects; the artifacts can be modeled by a polynomial, for example, and subtracted. The problem appears when we are interested in structures with a size comparable to the artifacts, such as the stellar halos of nearby galaxies.

The technique to overcome this difficulty is somewhat similar to the nodding technique used in near-infrared astronomy, where virtually all observations of astronomical sources are background dominated. The first key step is to take several images of the object of interest such that it is captured every time at a different location on the detector (we move the telescope by several arcmin between the individual frames). The background is calculated taking the median (in the simplest version) of all the individual frames  for every pixel. The resulting image does not show any object because a given pixel captures only blank sky in most of the images. The median image captures just the inhomogenities of the background caused by the parasitic light and flat fielding errors. The median image is then subtracted from the initial images, which makes their background perfectly flat. The background subtracted images are then stacked in the standard way to produce the final image. Obviously, the shifts between the individual frames should be larger than the object we intend to investigate. 

\section{The MATLAS survey}

\begin{figure}
	\includegraphics[width=0.99\textwidth]{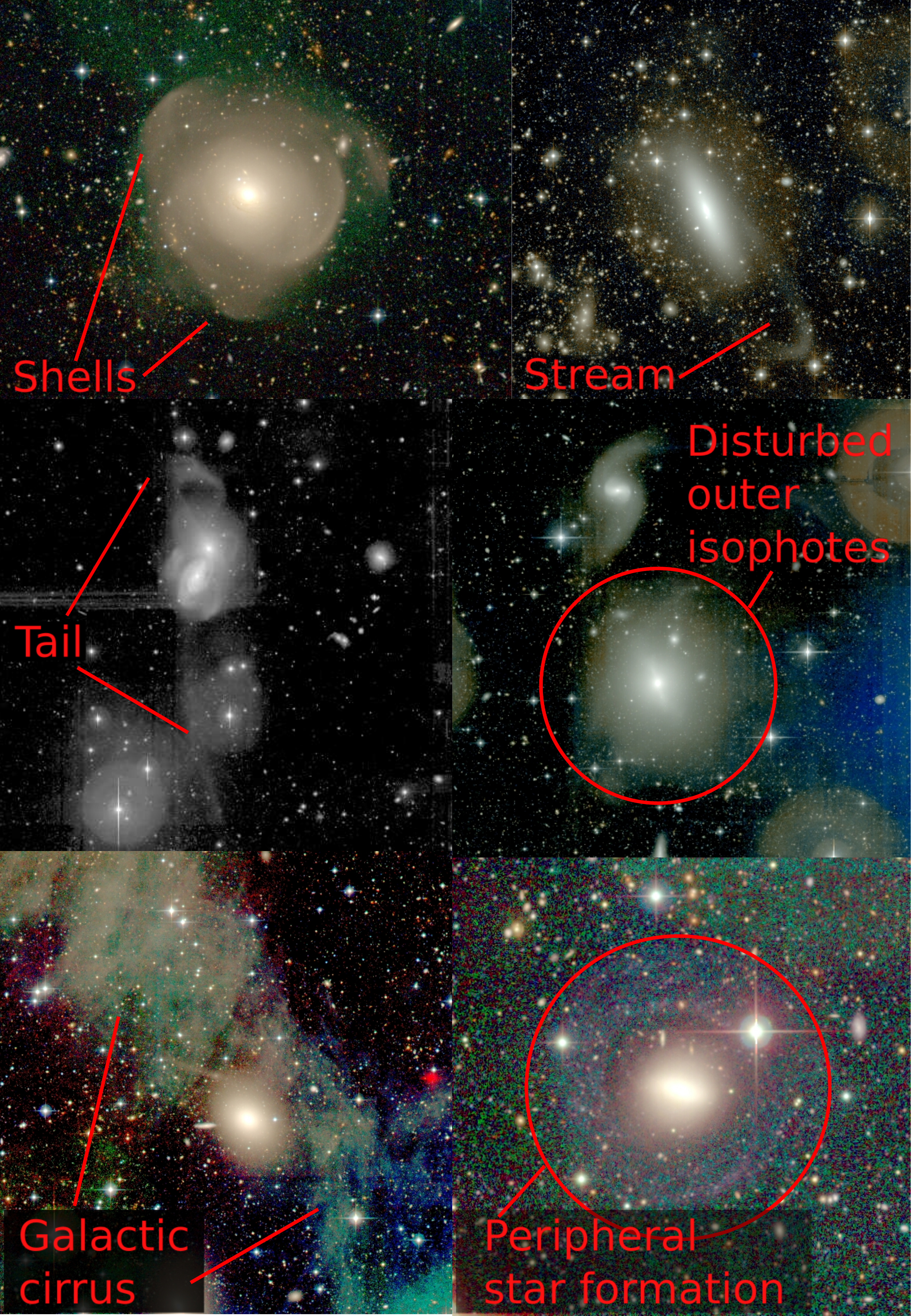}
	\caption{Examples of low-surface-brightness features detected in the MATLAS survey. The names of the galaxies are: NGC\,3619 (shells), NGC\,4684 (stream),  NGC\,3226 and NGC\,3227 (tidal tail), NGC\,3414 (disturbed outer isophotes),  NGC\,2592 (Galactic cirrus), UGC\,09519 (peripheral star formation). }
	\label{fig:ex}
\end{figure}

The MATLAS deep-imaging survey targets almost 200 nearby massive elliptical and lenticular galaxies. The images were taken with the MegaCam camera at the 3.6\,m Canada-France-Hawaii Telescope. Every galaxy was observed in between two and four photometric bands. The images reach the surface brightness limit of 28.5-29\,mag\,arcsec$^2$. The galaxies are very well explored since all of them belong to the sample of the ATLAS$^{3\mathrm{D}}$ project (Cappellari et al., 2011) that aims to collect as much information about its targets as possible. One of the main questions we wish to know about galaxies is how they were assembled. A galaxy can be assembled through two main channels. The first is the formation of stars directly in the galaxy. The second is merging of pre-existing galaxies. Merging means that two galaxies collide to form one bigger galaxy. The main purpose of MATLAS is to  look for signs of past galaxy mergers, the so-called tidal features. Examples of several types of them are shown in Figure~\ref{fig:ex} (shells, streams, tails, disturbed isophotes). The morphology of tidal features gives indications on the mass ratio and other properties of the galaxies that merged (see Wang et al, 2012 or B\'ilek et al., 2018 for some of the possible complications). The tidal features can remain visible several billions of years after  the two original galaxies have merged, even when hints of  the past collision have disappeared in the central merged body. The images in MATLAS are therefore highly useful in answering how typical a elliptical or lenticular galaxy forms. Fortunately, other objects also fall in the FOV of MegaCam  such as spiral galaxies, dwarfs or globular clusters, and may be studied thanks to the deep images, leading to unexpected discoveries. 

Once the images were taken, they were visually inspected and classified typically by 6 people per galaxy. 
The participants of our classification had at disposal a dedicated online image viewer. It is now available publicly at \url{http://obas-matlas.u-strasbg.fr}. In this navigator, we can  zoom in the image, change the brightness, contrast,  display the maps of the color index, or load the information on the objects  from the CDS/SIMBAD database and others. The participants were given a dedicated questionnaire with a list of various morphological structures and were asked to answer, for every galaxy, whether a given structure is present and if so, then how many of them are in the galaxy. More sophisticated annotation tools have also been developed.  

Figure~\ref{fig:ex} shows examples of the most important or interesting structures that we were looking for. {\bf Stellar shells} look like azimuthal sharp-edged glowing features. They form when two galaxies collide nearly radially. Recent observational studies suggested that shells can be formed from colliding galaxies of any mass ratio, but mergers of galaxies of very different masses are preferred  (Kado-Fong et al. 2018).

{\bf Stellar streams} appear as long thin structures. They usually form by disruption of a small galaxy by tidal forces. Stellar streams suggest the mergers of a mass ratio of about 1:4 or less. 

{\bf Tidal tails} which  have the same tidal origin as the streams, are often thicker and result from the interaction with a galaxy of a comparable mass or of a higher mass. They are more prominent if at least one of the interacting partners rotates.

% Figure (in PS or EPS format)

But all of these merger signs have a limited lifetime. When they get too old, the whole galaxy relaxes toward an equilibrium taking up a shape of a smooth ellipsoid or disk. Just before that, the signs of old mergers can be recognized  by the presence of  {\bf disturbed outer isophotes} in the galaxies. Some irregularities in the isophotes can also be induced by distant galaxy flybys. In the galaxy in the figure, both mechanisms probably acted.

The elliptical and lenticular galaxies in the MATLAS sample are characterized by their old stellar population. The deep images however revealed that some of the galaxies are surrounded by very faint disks of young stars. It seems that star formation in these galaxies was renewed after some period of inactivity. The reason for this {\bf peripheral star formation} is still unknown.

Finally, some of our images show {\bf galactic cirrus}. They are dust clouds in our own galaxy and their images are characteristic by their filamentary structure. In the context of the main goals of MATLAS they are considered as pollutants because they complicate the identification of the real structures in the galaxy of interest. On the other hand, MATLAS images are useful for investigating the properties of interstellar medium in our own galaxy (Miville-Desch{\^e}nes et al. 2016).  Incidentally, we found that the incidence of cirrus correlates excellently with the maps of extinction. Perhaps these images will help to construct more detailed maps of extinction.

The first results of the completed MATLAS survey have been published in the paper B\'ilek et al. (2020a). Its main goal was to provide the census of the various morphological structures.  In total, we found any signatures of galaxy merging in about 30-40 percent of galaxies. We also found that the frequency of such signs usually increases with the mass of the galaxy and the number of its neighbors. This is expected: galaxies encounter more frequently in denser environments and more massive galaxies are more effective in accreting other galaxies because of their stronger gravitational fields. The comparison of the observations to theoretical expectations is planned. In the future, we also want to make use of our results for investigating the formation of galaxies from a purely observational point of view.  It will be possible, for example, to look for the connection of galaxy interactions with various peculiar properties of galaxies (e.g., kinematically distinct cores, active galactic nuclei, etc.).

\section{Deep imaging at Milankovi\'c Telescope}

We tested the observing methods similar to those used for MATLAS using the new 1.4\,m Milankovi\'c Telescope in Serbia. We found that it can be used for deep imaging very successfully. For example we found that the tip the faint stream in the northern part of NGC\,474 that is barely visible in the MATLAS image  can be captured with Milankovi\'c in about 3-4 hours. The broadband $L$ filter has to be used for that (it covers most of the optical part of the spectrum). Ana Vudragovi\'c gives more details in this book of proceedings. The downsides of deep imaging with Milankovi\'c instead of  MegaCam at CFHT are the substantially smaller field of view ($13.3^\prime\times13.3^\prime$ vs. $60^\prime\times60^\prime$), higher seeing FWHM (typically $1-2^{\prime\prime}$ vs. $<1^{\prime\prime}$), and longer observing times are needed when using filters that are narrower than the $L$ filter (many hours vs. 40\,min). Actually these are not serious disadvantages if suitable targets are chosen. The advantage, compared to MegaCam is a lower scattering of light on the optics of the telescope (less light in the wings of the point spread function)  and the point spread function is smoother. This will allow modeling the parasitic glow around bright stars easily and subtracting it to visualize the faint structures (work in progress).

Our first published deep-imaging application of Milankovi\'c was the observation of the stellar stream at the galaxy NGC\,5907. Initially, amateur astronomers (e.g., Mart{\'\i}nez-Delgado et al. 2008) detected repeatedly a stream that forms a double loop wrapping the galaxy. It was a big surprise when  professional astronomers equipped by a telescope optimized for deep imaging reported that the stream has a morphology of an arc (van Dokkum et al. 2019). Our deep image agreed with the morphology of the stream seen by the professionals (M{\"u}ller et al. 2019). The reason why the new observations are different is not clear yet.

We made use of deep imaging also in  B{\'\i}lek et al. (2020b). Here we aimed to reveal more about the nature of an enigmatic ``dark cloud'' in the Virgo cluster of galaxies. Several of such objects were revealed by a blind HI survey (Taylor et al. 2012). These dark clouds are unique and difficult to interpret because their wide HI lines combine with large distances of the clouds from the nearest neighbor galaxies. The clouds were never captured optically. There is a hypothesis that they are dark matter halos containing gas but little or no stars (Taylor et al., 2016, but other possibilities exist, such as that they are tidal debris, Duc \& Bournaud, 2008). We took a very deep image of one of these objects in the hope to detect a faint galaxy. The image did not show it. This allowed us to estimate that the luminosity of the object is below $1.1\times10^7$ of the solar and that the object contains at least 3 times more mass in gas than in stars. This is another illustration of  the potential of deep imaging.

ACKNOWLEDGEMENTS: We thank the anonymous referee for very useful comments. MB acknowledges the support by Cercle Gutenberg.

% References
\references
B{\'\i}lek, M., Thies, I., Kroupa, P., Famaey, B. : 2018, \journal{Astron. Astrophys.}, 614, A59.

B{\'\i}lek, M., Duc, P.-A., Cuillandre, J.-C., Gwyn, S., Cappellari, M., Bekaert, D.~V., Bonfini, P., et al. : 2020a, \journal{Mon. Not. R. Astron. Soc.}, 498, 2138. 

B{\'\i}lek, M., M{\"u}ller, O., Vudragovi{\'c}, A., Taylor, R. : 2020b,  \journal{Astron. Astrophys.}, 642, L10. 
 
Cappellari, M., Emsellem, E., Krajnovi{\'c}, D., et al. : 2011, \journal{Mon. Not. R. Astron. Soc.}, 413, 813. 

Duc P.-A., Bournaud F. : 2008, \journal{Astrophys. J.}, 673, 787.

Duc, P.-A., Cuillandre, J.-C., Karabal, E., Cappellari, M., Alatalo, K., Blitz, L., Bournaud, F., et al. : 2015, \journal{Mon. Not. R. Astron. Soc.}, 446, 120.

Kado-Fong, E., Greene, J.~E., Hendel, D., Price-Whelan, A.~M., Greco, J.~P., Goulding, A.~D., Huang, S., et al. : 2018, \journal{Astrophys. J.}, 866, 103. 

 Mart{\'\i}nez-Delgado, D., Pe{\~n}arrubia, J., Gabany, R.~J., Trujillo, I., Majewski, S.~R., Pohlen, M. : 2008, \journal{Astrophys. J.}, 689, 184.

Miville-Desch{\^e}nes, M.-A., Duc, P.-A., Marleau, F., Cuillandre, J.-C., Didelon, P., Gwyn, S., Karabal, E. : 2016, \journal{Astron. Astrophys.}, 593, A4.

 M{\"u}ller, O., Vudragovi{\'c}, A., B{\'\i}lek M., 2019, \journal{Astron. Astrophys.}, 632, L13.

Taylor, R., Davies, J.~I., Auld, R., Minchin, R.~F. : 2012,  \journal{Mon. Not. R. Astron. Soc.}, 423, 787. 

Taylor, R., Davies, J.~I., J{\'a}chym, P., Keenan, O., Minchin, R.~F., Palou{\v{s}}, J., Smith, R., et al., 2016, \journal{Mon. Not. R. Astron. Soc.}, 461, 3001.

van Dokkum, P., Gilhuly, C., Bonaca, A., Merritt, A., Danieli, S., Lokhorst, D., Abraham, R., et al. : 2019, \journal{Astrophys. J.}, 883, L32. 

Wang, J., Hammer, F., Athanassoula ,E., Puech, M., Yang, Y., Flores, H. : 2012,  \journal{Astron. Astrophys.}, 538, A121.

\endreferences

\end{document}